\documentclass{ifacconf}

\usepackage{graphicx}      
\usepackage{natbib}        
\usepackage{amsmath,amssymb,amsfonts}
\usepackage{amsmath}
\usepackage{subcaption}
\usepackage{caption}
\usepackage{stfloats}
\usepackage{abraces}

\usepackage{algorithm}
\usepackage{algorithmic}
\usepackage{upgreek}
\usepackage{tikz}
\usetikzlibrary{arrows.meta}   
\usepackage{circuitikz}
\usepackage{graphicx}          
\usepackage{amsmath,amssymb}   

\def\begmat#1{\begin{bmatrix}#1\end{bmatrix}}

\def\cale{{\cal E}}

\def\calc{{\cal C}}

\def\calb{{\cal B}}

\def\cals{{\cal S}}
\def\cale{{\cal E}}

\def\cala{{\cal A}}

\def\L2e{{\cal L}_{2e}}

\def\rea{\mathbb{R}}

\def\diag{\mbox{diag}}

\def\col{\mbox{col}}

\def\diag{\mbox{diag}}
\def\rank{\mbox{rank}\;}

\def\min{{\mbox{min}}}
\def\max{{\mbox{max}}}

\usepackage{color}

\newtheorem{problem}{\bf Problem}

\usepackage{algorithm}
\usepackage{xcolor}
\definecolor{codegreen}{RGB}{0,100,0}

\begin{document}
\begin{frontmatter}

\title{Data-Driven Adaptive Output Regulation
of Unknown Linear Systems \thanksref{footnoteinfo}} 

\thanks[footnoteinfo]{
This work was supported in part by the National Natural Science Foundation of China under Grant 62573383 and by Natural Sciences and Engineering Research Council of Canada (NSERC) under Grant RGPIN-2024-0478, and the Programme PIED. }

\author[zju]{Shangkun Liu}, 
\author[zju]{Lei Wang},
\author[poly]{Bowen Yi}

\address[zju]{College of Control Science and Engineering, Zhejiang University,
Hangzhou 310027, China
}
\address[poly]{Department of Electrical Engineering, Polytechnique Montr\'eal \& GERAD, Montr\'eal, Canada}
\address{\rm E-mail: \texttt{\{shangkun.liu,lei.wangzju\}@zju.edu.cn, bowen.yi@polymtl.ca}}

\begin{abstract}                
This paper investigates the linear output regulation problem with both the exosystem and the plant fully unknown. A data-driven regulator is proposed to achieve asymptotic regulation and closed-loop stability without performing model identification. The method constructs a nominal approximate internal model and filters of input and outputs, thereby yielding a stabilizable cascaded nominal system whose states are available. For this nominal system, a stabilizing law is derived from an offline dataset that has been acquired from the plant during experiments, such that the system states exponentially converge to a subspace. An identifier in discrete-time is, then, implemented to correct the internal model and update the stabilizing law; as a result, the regulation error can be steered to zero asymptotically under some persistent excitation conditions.  
\end{abstract}




\end{frontmatter}
\vspace{-2em}
\section{Introduction}
\label{sec:1}
\vspace{-1.2em}



The output regulation problem aims to design a controller that enables the system output to track a reference signal or reject disturbances generated by an autonomous exosystem. A seminal regulator design framework has been proposed by introducing a certain copy of the exosystem dynamics to compensate for the effect of exosystem, i.e., the internal model principle \cite{francis1975internal,francis1976internal}. In recent decades, such an internal model-based framework has  attracted extensive attention in the control community, particularly for nonlinear systems ranging from single-input single-output (SISO) systems \cite{Isi1990out,Huang1990on,huang2004nonlinear}, to multi-input multi-output (MIMO) \cite{wang2016nonlinear,wang2020pre,wang2020robust} and multi-agent \cite{isidori2017lectures,su2011cooperative} scenarios. 

In the output-regulation problem, a conventional assumption is that the steady-state control input satisfies an immersion condition \cite{huang2004nonlinear}, which requires a priori knowledge of the exogenous dynamics and imposes  structural conditions on the plant. To handle such issue, \cite{marconi2007output,marconi2008uniform} propose a Luenberger observer-based design scheme. However, the design of the corresponding steady-state compensator of control input turns out non-trivial. Subsequent works then have been developed by proposing identification-based approximate/adaptive design schemes \cite{bin2019adaptive}. Besides, several other notable adaptive internal model regulators have also been proposed in \cite{wang2019adaptive,priscoli2006adaptive,serrani2002semi}. These contributions predominantly address minimum-phase systems, while very few results treat adaptive regulation issues for non-minimum-phase plants, with the exception of \cite{bin2019adaptive}. In \cite{bin2019adaptive} a general linear plant with an unknown exosystem is considered, and a discrete-time identifier is designed to handle internal-model parameter identification. Nevertheless, how to extend this result to accommodate uncertainties in the plant itself remains an open problem.

Recently, data-driven control has emerged as a powerful paradigm for scenarios in which the plant model is unknown. \cite{berberich2021data} and \cite{bosso2025derivative} extended these ideas to continuous-time linear systems and proposed a suite of solutions for the case of unknown plant models. Building on results in \cite{HUetal,bosso2025data}, this approach has been further developed for output regulation by designing controllers directly from data, which has given rise to cascade-based design methods such as those in \cite{Mao,mao20252}. However, in the most of the foregoing data-driven output-regulation works the exosystem is assumed known; the more general and challenging scenario where both the exosystem and the plant structure are unknown remains open.

To bridge this gap, this paper develops a data-driven adaptive output regulator for linear systems in which both the plant dynamics and the exosystem are fully unknown. A stabilizable nominal cascaded structure is constructed from measured input–output data, within which an adaptive internal model with online parameter updates is embedded. Building on this structure, a data-driven stabilizing law together with an adaptive mechanism is designed to achieve asymptotic tracking and ensure closed-loop stability, all without identifying either the plant or the exosystem model.

\emph{Notation.} $|\cdot|$ denotes the Euclidean norm and $\dagger$ denotes the Moore–Penrose pseudoinverse. 
{We use $\mathbb{C}$ to represent the complex plane, and $\mathbb{C}_{+}$ for the closed right half-plane.} For square matrices $A$ and $B$, the notion $A \prec B$ (or $A\preceq B$) indicates that $(A-B)$ is negative definite (or semidefinite). We use $\rea^{m\times m}_{\succ 0}$ (or $\rea^{m\times m}_{\succeq 0}$) to represent the set of $m\times m$ positive definite (or semidefinite) matrices. $\sigma(A)$ denotes the spectrum of the matrix $A  \in \rea^{n\times n}$. When clear from the context, the arguments of mappings and operators are omitted. We use $\tilde{y}(t)$ to represent the measured data corresponding to the true state $y(t)$ at time $t$. For a vector $\Xi = \col(x_1, \ldots, x_m) \in \rea^{nm}$, where $x_i \in \rea^n$, we define the reshape operator $\gamma(\Xi) := [x_1~\ldots ~ x_m] \in \rea^{n \times m}$, i.e. $\mbox{vec}(\gamma(\Xi)) = \Xi$.

\section{Problem Formulation \& Preliminary}
\label{sec:2}

\subsection{Problem Formulation}
Consider the SISO linear time-invariant (LTI) system given by
\begin{equation}\label{1-eq-sys}
\dot{x}  =  Ax+Bu, \quad 
y  =  C x,
\end{equation}
{where $x\in \rea^n$ is the state,  {$u\in\rea$ is the control input, $y \in \rea$ is the measured output. The matrix triplet $(A,B,C)\in\mathbb{R}^{n\times n}\times \mathbb{R}^{n}\times \mathbb{R}^{1\times n}$ is \emph{unknown} except the dimension $n\in \mathbb{N}_+$, and satisfies the following.

\begin{assum}\label{ass1}
The pair $(A, B)$ is controllable, and the pair $(C,A)$ is observable. 
\end{assum}\par

In this paper, we are interested in the output regulation problem, where we design a controller such that the output $y$ asymptotically converges to the reference output $y_r \in \rea$ that is generated by an exogenous system
\begin{equation}\label{2-eq-exosys}
\dot{w} = Sw, \quad 
{y}_r  =  C_r w,
\end{equation}
with $w \in \rea^{d}$ the exogenous system state. The pair $(C_r, S)$ is \emph{unknown} in the controller synthesis but assumed to be observable. Additionally, the eigenvalues of $S$ are distinct on the imaginary axis, thus $w$ being bounded over time. 

We further impose the following assumptions.

\begin{assum}\label{ass2} 
There exists a known compact set $\mathcal{E} \subset \mathbb{R}^d $ such that for any $\theta:=\col(\theta_1,\ldots,\theta_d)\in\mathcal{E}$, there holds
\begin{equation*}\begin{aligned}\mathrm{rank}
\begin{bmatrix}
A-\lambda I_n & B \\
C & 0
\end{bmatrix}= n + 1 \,
\end{aligned}\end{equation*}
for all roots $\lambda$ of the characteristic polynomial 
$
{p}(\lambda,{\theta})=\lambda^d + \theta_d\lambda^{d-1}  \ldots + \theta_2\lambda + \theta_1. 
$
Moreover, {the characteristic polynomial of $S$ is ${p}(\lambda,{\theta}^\star)$ for some (unknown) $\theta^\star \in\mathcal{E}$. }
\end{assum}

\begin{assum}\label{ass3}
{The initial condition $w(0)$ guarantees that $w(t)$ lies within a known compact set $W \subset \mathbb{R}^{d}$ for all $t \ge 0$.}
\end{assum}\par

{In this paper, we are interested in the following problem.}

\begin{problem}
\label{problem:1}
Suppose that the input-output dataset 
\begin{equation*}
  \cals_\star :=\{ \tilde u(t), \tilde y(t)\}_{[0, t_\star]}
\end{equation*}
has been collected from the plant \eqref{1-eq-sys} during a prior experiment of duration $t_\star>0$, where $\tilde u(\cdot)$ and $\tilde y(\cdot)$ denote the recorded input and output trajectories over $[0, t_\star]$.
The objective is to design a controller that leverages the offline dataset $\cals_\star$, together with the online measurements $(u,y,y_r)$ obtained from systems \eqref{1-eq-sys}-\eqref{2-eq-exosys}. The matrices $(A,B,C,S,C_r)$ are assumed to be unknown, except for the state dimension $n$ and the exosystem order $d$. The controller needs to ensure that all closed-loop states are bounded and that the tracking error $e:= y- y_r$ satisfies
\begin{equation}
\label{eq:e_conv}
\lim_{t\to\infty} e(t)=0.
\end{equation}
\end{problem}

\begin{rem}
    An intuitive approach to achieve the desired regulation objective relies on transforming \eqref{2-eq-exosys} into an observable canonical form and identifying the matrix $S$ from $y_r$. However, when  $y_r$ is not directly available, e.g., in multi-agent scenarios where \eqref{2-eq-exosys} serves as the leader and most followers do not have access to $y_r$, such a method becomes infeasible. The approach proposed here avoids explicit identification of $S$ and is therefore particularly well suited for multi-agent scenarios.
\end{rem}

\subsection{Preliminaries}

For output regulation, we adopt the ``approximate" internal model
\begin{equation}\label{4-eq:eta}
\begin{aligned}
&\dot{\eta}=\Phi(\theta)\eta+Ge,\end{aligned}
\end{equation}
with matrices
\begin{equation}\label{5-PhiG}
\begin{aligned}
\Phi :=
\begin{bmatrix}
0 & 1 \\
\vdots&& \ddots \\
0&&&1\\
-\theta_1 & -\theta_2 & \ldots & -\theta_{d}
\end{bmatrix} \in \rea^{d\times d}
,
\;
G = \begmat{0 \\ \vdots \\ 0 \\ 1} \in \rea^{d}\,
\end{aligned}
\end{equation}
where $\Phi$ is parameterized by a vector $\theta\in\mathcal{E}$. 
If the exosystem matrix $S$ is known, one can set $\theta = \theta^\star$ such that $S$ has the same characteristic polynomial as $p(\lambda,\theta^{\star})$. When the system matrices $(A,B,C)$ are also known, this problem is fully addressed by the internal model principle \cite{francis1976internal}.

For the case where the system \eqref{1-eq-sys} is unknown but the exosystem matrix $S$ is available, a data-driven control method was recently proposed in \cite{bosso2025data}. In contrast, Problem \ref{problem:1} deals with an unknown matrix $S$, which represents a more practical yet more challenging scenario. Before presenting our approach, we briefly recall the key results from \cite{bosso2025data}, in which a linear filter is constructed for the input $u$ and the output $y$, i.e. 
\begin{equation}
\label{6-eq:filter}
\begin{aligned}
\dot{{\zeta}}_y  = F {\zeta}_y + Ly,\quad \dot{{\zeta}}_u = F {\zeta}_u + L u.
\end{aligned}
\end{equation}
Here, we select a Hurwitz matrix $F \in \mathbb{R}^{n\times n}$ with distinct eigenvalues $\lambda_1, \ldots, \lambda_n$ and a vector $L \in \mathbb{R}^{n}$ such that $(F,L)$ is controllable. It has shown the following.

\begin{lem}\label{lem4}
    Let Assumption \ref{ass1} hold. There exist matrices $(\Pi_1,H_1)$ and $(\Pi_2,H_2) \in \mathbb{R}^{n\times n} \times \mathbb{R}^{1\times n}$ satisfying
\begin{equation}
\begin{aligned}\label{7-pi1}
\Pi_1F  &= (A- \Pi_1LC) \Pi_1,\quad
H_1 = C \Pi_1 ,\\
\Pi_2F  &= (A- \Pi_1LC) \Pi_2,\quad \Pi_2 L  = B,\quad H_2 = C \Pi_2,
\end{aligned}
\end{equation}
where $[\Pi_1\; \Pi_2]$ has full row rank. Additionally, for matrices
\begin{equation}
\label{8-eq:ABC}
    \mathcal{A} = \begin{bmatrix}
        F+LH_1 & LH_2\\ 0 & F
    \end{bmatrix},
    \;
\mathcal{B} = \begin{bmatrix}
        0\\  L
    \end{bmatrix},
    \;
\mathcal{C} = \begin{bmatrix}
        H_1 & H_2
    \end{bmatrix},
\end{equation}
we can then establish the following properties.
\begin{enumerate}
    \item[P1:] The pair $(\mathcal{A},\mathcal{B})$ is stabilizable, and $(\mathcal{C}, \mathcal{A})$ detectable;
    \item[P2:] For any $\theta\in\mathcal{E}$, there holds
\begin{equation}\label{9-eq:nonreson} \begin{aligned}\mathrm{rank}
\begin{bmatrix}
\mathcal{A}-\lambda I_{2n} & \mathcal{B} \\
\mathcal{C} & 0
\end{bmatrix}= 2n + 1 , \quad \forall\lambda\in\sigma(\Phi(\theta)).
\end{aligned}\end{equation}
\end{enumerate}
\end{lem}

An important observation from the above lemma is that the system \eqref{7-pi1}, together with the output 
 $$
 y = H_1 {\zeta}_y + H_2 {\zeta}_u,
 $$
 is a non-minimal realization of \eqref{1-eq-sys}. To be precise, 
$$
\begin{aligned}
    x(0)= \Pi_1 {\zeta}_y(0) + \Pi_2 {\zeta}_u(0) 
    \;
    \Rightarrow
    x(t) = \Pi_1 {\zeta}_y(t) + \Pi_2 {\zeta}_u(t)
,
    \forall t.
\end{aligned}
$$
When implementing the filter~\eqref{7-pi1}, although the initial conditions 
$(\zeta_y(0), \zeta_u(0))$ cannot be properly selected, it is evident that the transverse 
coordinate 
\begin{equation*}
\rho := x - \Pi_1 {\zeta}_y - \Pi_2 {\zeta}_u
\end{equation*}
is exponentially stable. Its dynamics are given by
\begin{equation*}
   \dot{\rho} = (A - \Pi_1 L C)\rho, 
\end{equation*}
where $(A - \Pi_1 L C)$ is, similar to $F$, Hurwitz. That being said, the plant \eqref{1-eq-sys} can be exactly represented by the dynamics of $({\zeta}_y, {\zeta}_u, \rho )$ in order to overcome the unavailability of the signal $x(t)$. As a consequence, we translate Problem \ref{problem:1} into the output regulation of the cascaded system composed of the filter \eqref{6-eq:filter}, the internal model, and the exogenous system \eqref{2-eq-exosys}. 



\section{The regulator structure}
\label{sec:3}

In this section, we propose our approach that is independent of the unknown signal $w$, consisting of a post-processing internal model, an online parameter identifier for $\theta^\star$, and a filter. The internal model guarantees that the regulation error $e$ converges to zero; the identifier updates the inexact parameter $\theta$ using the discrete mechanism in \cite{bin2019adaptive}; and the filter estimates the unknown states $x$, thus providing reliable data for controller computation. 

\subsection{Nominal Approximate Internal Model}

Instead of using \eqref{4-eq:eta}, we consider the following internal model  
\begin{equation}
\label{9-eq:etay-nom}
\dot{\eta}_y=\Phi(\theta)\eta_y + Gy,
\end{equation}
which is, indeed, a nominal formulation independent of the exogenous signal $w$. Together with the filter \eqref{6-eq:filter} and the algebraic relation in Lemma \ref{lem4}, the dynamics of $(\eta_y, x)$ can be represented by the non-minimal realization
\begin{equation}
\label{10-eq:supersysw-nom}
\begmat{\dot{\eta}_y \\ \dot{\zeta}_y \\ \dot{\zeta}_u \\ \dot \rho}
= 
\begmat{\Phi(\theta) & GH_1 & GH_2 & GC
\\
0 & F+ LH_1 & L H_2 & LC
\\
0 & 0 & F & 0
\\
0 & 0 & 0 & A-\Pi_1LC
}
\begmat{\eta_y \\ \zeta_y \\ \zeta_u \\ \rho}
+
\begmat{0 \\ 0 \\  L \\ 0} u.
\end{equation}
The above system satisfies the following.

\begin{lem}\label{lem6}
    With Assumptions \ref{ass1}-\ref{ass2}, the system \eqref{10-eq:supersysw-nom} is stabilizable for any $\theta\in\mathcal{E}$.
\end{lem}
\begin{pf}
In terms of the structure of the system matrix, the state $\rho$ is within the uncontrollable subsystem. Since the matrix $(A-\Pi_1 LC)$ is Hurwitz, the stabilizability of equivalent to the one of the point pair
    $$
    \left(
    \begmat{\Phi(\theta) &  G \mathcal{C}\\ 0 & \cala  }
    ,
    \begmat{0 \\ \calb}
    \right)
    $$
    with $(\cala,\calb,\calc)$ defined in \eqref{8-eq:ABC}. Applying the PBH test, this is equivalent to for any $\theta \in \cale$,
\begin{equation}\label{11-rowrankmatx}
\begin{aligned}
\rank
\begin{bmatrix}
 \Phi(\theta)-\lambda I_d & G\mathcal{C} &0 \\
 0 & \mathcal{A}-\lambda I_{2n } & \mathcal{B} \\
\end{bmatrix}
=
d+ 2n, 
\; \forall \lambda \in \mathbb{C}_+.
\end{aligned}
\end{equation}
To facilitate the analysis, we transform the column of matrix \eqref{11-rowrankmatx} corresponding to $\Phi(\theta)$ by right-multiplying a lower triangular Toeplitz matrix $T_{\Phi_{\text{right}}} = [\,\lambda^{i-j}\,]_{i,j=1}^d,$ with $  \lambda^{i-j}=0\ \text{for } i<j.$
As a result, the condition \eqref{11-rowrankmatx} is equivalent to for all $\lambda \in \mathbb{C}_+$
\begin{equation}\label{12-rankmatrix}
\begin{aligned}
\rank
\begin{bmatrix}
 \begin{bmatrix}I_{d-1} & 0 \\
0 & -{p}(\lambda,\theta)
\end{bmatrix}  & \begin{matrix}0 \\ \mathcal{C} \end{matrix}
  & \begin{matrix}0 \\ 0 \end{matrix} \\
 0 &  \mathcal{A} - \lambda I_{2n } &  \mathcal{B} 
\end{bmatrix}
= 
d+2n,
\end{aligned}
\end{equation}
where the definition of $G$ in \eqref{5-PhiG} is used. 

For $\lambda \in \mathbb{C}_+\setminus\sigma(\Phi(\theta))$, we have ${p}(\lambda,\theta) \neq 0$. Then matrix \eqref{12-rankmatrix} is full-row rank if and only if there holds $\rank \begin{bmatrix}
 \mathcal{A} - \lambda I_{2n } &  \mathcal{B} 
\end{bmatrix} = 2n\,.$ This is true by recalling the Property P1 that the pair $(\mathcal{A},\mathcal{B})$ is stabilizable. For $\lambda \in \sigma(\Phi(\theta))$, 
matrix \eqref{12-rankmatrix} is full-row rank if and only if there holds the non-resonance condition \eqref{9-eq:nonreson}.
\hfill $\square$
\end{pf}

From the above, there exists a linear feedback that exponentially stabilizes  system \eqref{10-eq:supersysw-nom}. For convenience, we write
$$
\zeta_{oi} := \col(\zeta_y, \zeta_u).
$$
However, although the signals $(\eta_y, \zeta_{oi})$ are available for controller design—whether model-based or data-driven—the signal $\rho$ is not accessible. To this end, similarly to \cite{bosso2025data}, we perform a change of coordinate 
$$
(\eta_y, \zeta_{oi}, \rho ) \mapsto (\eta_y, \zeta_{oi}, T_\rho \rho)
$$
with the nonsingular matrix $T_\rho \in \rea^{n\times n}$ satisfying
$
    \Lambda_F = T_\rho (A-\Pi_1LC) T_\rho^{-1}
$
and $\Lambda_F := \diag(\lambda_1, \ldots, \lambda_n)$.
It always exists since $(A-\Pi_1L C)$ is similar to $F$, equivalently to $\Lambda_F$, and ${d\over dt}(T_\rho \rho) = \Lambda_F (T_\rho \rho)$. Therefore, we use the following dynamics
\begin{equation}
\label{13-dot:chi}
    \dot{ \chi} = \Lambda_F \chi, \quad \chi (0)= \mathbf{1}_n,
\end{equation}
and there exists a matrix $M_\rho \in \mathbb{R}^{1\times n}$ such that $C \rho = M_\rho \chi$. The stabilizable plant \eqref{10-eq:supersysw-nom} can be rewritten as
\begin{equation}\label{14-chisys}
\begmat{\dot{\eta}_y \\ \dot{\zeta}_{oi} \\ \dot {\chi} \\ }
=
\begmat{\Phi(\theta) & G\calc & G M_\rho \\ 0 & \cala  & \begmat{LM_\rho \\ 0}\\0 & 0 & \Lambda_F}
\begmat{\eta_y \\  \zeta_{oi} \\ \chi} + \begmat{0 \\ \calb \\ 0}u.
\end{equation}
This system is also stabilizable, but all the signals $(\eta_y, \zeta_{oi}, \chi)$, along with their time derivatives, are available in both online and offline contexts.

\subsection{Post-processing Data} 
For the continuous-time signals in the dataset $\cals_\star$, we sample their values at times $t_0, \ldots, t_k$ and define the data matrices accordingly
$$
\begin{aligned}
Y &:=
\begin{bmatrix}
 \tilde y(t_{0}) &  \ldots &  \tilde y(t_{k})
\end{bmatrix},
\quad 
  U:=
\begin{bmatrix}
\tilde u(t_{0}) &  \ldots & \tilde u(t_{k})
\end{bmatrix}.
\end{aligned}
$$
Now we post-process the dataset $\cals_\star$. To be precise, we simulate the filters \eqref{6-eq:filter} and \eqref{13-dot:chi} and the nominal internal model \eqref{9-eq:etay-nom} by injecting $\tilde u (\cdot)$ and $\tilde y(\cdot)$ over $[0, t_\star]$. Accordingly, we obtain the ``post-processed'' dataset
$$
\cals_p^\theta:=\{ \tilde \eta_y^\theta(t), \tilde \zeta_{oi}(t),   \tilde \chi(t)\}_{[0, t_\star]}.
$$
Note that we add the superscript $(\cdot)^\theta$ to indicate that the post-processed signal $\eta_y$ is dependent on the parameter $\theta$ in the nominal internal model.

Similarly, we sample the dataset $S_p^\theta$ and obtain the following data matrices:
$$
\begin{aligned}
Z^\theta_{\eta,+} &:= 
\begmat{\dot{\tilde\eta}^\theta_y(t_0) & \ldots &  \dot{\tilde{\eta}}^\theta_y(t_k)}
,
\quad
Z^\theta_{\eta} :=
\begmat{\tilde \eta^\theta_y(t_0)  & \ldots  & \tilde\eta^\theta_y(t_k)},
\\
Z_{\zeta_{oi},+} & :=
\begmat{{\dot{\tilde \zeta}_{oi}}(t_0) & \ldots & \dot{\tilde \zeta}_{oi}(t_k)},
Z_{\zeta_{oi}} :=
\begmat{{\tilde \zeta}_{oi}(t_0) & \ldots & {\tilde \zeta}_{oi}(t_k)},
\\
X & :=
\begmat{\tilde {\chi}(t_0) & \ldots & \tilde{\chi}(t_k)}.
\end{aligned}
$$
Note that the superscript in the matrices $Z_{\eta,+}^\theta$ and $Z_{\eta}^\theta$ is included to emphasize their dependence on $\theta$ in the internal model.

Obviously, for a fixed $\theta$, the above matrices satisfy
\begin{equation}\label{15-datasys}
 Z^\theta_+ = \mathcal{A}_c^\theta Z^\theta + \mathcal{B}_c U + \mathcal{D}{M}_{\rho} X, 
\end{equation}
with $Z^\theta_+=\col(Z^\theta_{\eta,+},Z_{\zeta_{oi},+}),
\;
Z^\theta=\col(Z^\theta_{\eta},Z_{\zeta_{oi}})$, where 
\begin{equation*}
 \mathcal{A}_c^\theta := \begin{bmatrix}
  \Phi(\theta) & G\mathcal{C} \\ 0 & \mathcal{A}
\end{bmatrix}
,\;
\mathcal{B}_c=\begin{bmatrix}
        0 \\ \mathcal{B}
    \end{bmatrix}
,\;
\mathcal{D}=\begmat{G \\ L \\ 0_n}.  
\end{equation*}
Note that the matrices $(H_1,H_2,M_\rho)$ are unknown. We make the important observation that the matrices $(Z_{\zeta_{oi}}, X)$ do not rely on the parameter $\theta$, and thus we impose the following excitation condition on the post-processed data.
\begin{assum}
\label{ass:7}
For the sampling instances $\{t_0,\ldots, t_k\}$, the post-processed dataset satisfies the excitation condition
\begin{equation*}
    \mbox{rank}\begmat{Z_{\zeta_{oi}}\\ X}=3n.
\end{equation*}
\end{assum}

\begin{lem}\label{lem8}
Let Assumptions \ref{ass1}, \ref{ass2}  and \ref{ass:7} hold. Then, there exists $P\in \mathbb{R}^{(2n+d)\times(2n+d)}_{\succ 0}$ such that
\begin{equation}
\label{eq:LMI7}
\begin{aligned}
&\begin{bmatrix}
\hat{\mathcal{A}}^{\theta\top} P+P\hat{\mathcal{A^{\theta}}}+Q & P\mathcal{B}_c\\
\mathcal{B}_c^\top P & R
\end{bmatrix}\preceq 0 ,\end{aligned}
\end{equation}
for some $Q \in \mathbb{R}^{(2n+d)\times(2n+d)}_{\succeq 0}$ and $R\in\mathbb{R}_{>0}$, where
\begin{equation*}
\begin{aligned}
\hat{\mathcal{A}^{\theta}}&=\mathcal{A}^{\theta}_n +  \left(Z^\theta_+ - \mathcal{A}^{\theta}_nZ^\theta - \mathcal{B}_c U \right)
\begin{bmatrix}
Z_{\zeta_{oi}}\\  X
\end{bmatrix} ^{\dagger} 
\begin{bmatrix}
I_{2n }\\  0_{d\times 2n
}\end{bmatrix}
\begin{bmatrix}0_{2n} & 1 \end{bmatrix},
\\
\mathcal{A}^{\theta}_n& =\diag({\Phi(\theta), I_2\otimes F}).
\end{aligned} 
\end{equation*}
The dynamic output-feedback \eqref{6-eq:filter}, \eqref{9-eq:etay-nom}, \eqref{13-dot:chi}, and
\begin{equation}\label{39-u=kz}
    u = K^{\theta} \begmat{\eta_y \\ \zeta_{oi}}
\end{equation}
with $ K^{\theta} = R^{-1}\mathcal{B}_c^\top P$ globally exponentially stabilize the system \eqref{10-eq:supersysw-nom} at the origin.
\end{lem}\par


We have the following.

\begin{thm}\label{thm1}
Let $K^\theta$ be such that the state feedback law \eqref{39-u=kz} globally exponentially stabilizes system \eqref{10-eq:supersysw-nom} at the origin, and consider the filter \eqref{6-eq:filter} and
\begin{equation}\label{17-filter}
\begin{aligned}
    \dot \eta_e & = \Phi(\theta) \eta_e + Ge
    \\
    \dot \zeta_r & = F \zeta_r + L y_r.
\end{aligned}
\end{equation}
Then, the subspace
\begin{equation}
\label{18-eq:M}
\mathcal{M} = \{ \eta_e, \zeta_u, \zeta_y, \zeta_r, w: \col(\eta_e,\zeta_y - \zeta_r, \zeta_u ) = \Psi w \}    
\end{equation}
is globally exponentially stable (GES)\footnote{This is in the sense of global exponential stability with respect to a closed invariant set; see, for example, \cite[Def. 2]{LINetal}.} for any $\theta \in \mathcal{E}$, for the system \eqref{17-filter}, where $\Psi=\col( \Psi_{\eta_e} , \Psi_{\zeta_y - \zeta_r} , \Psi_{\zeta_u})\in \mathbb{R}^{(d+2n) \times d}$ and obtaining
\begin{equation}\label{19-Phi-CL}
\begin{aligned}
\Psi_{{\eta}_{ei}} S &= \Psi_{{\eta}_{e(i+1)}}\,,\quad i=1,\ldots,d-1, \\
\Psi_{{\eta}_{ed}} S &=  -\theta_{1} \Psi_{{\eta}_{e1}} -\theta_2 \Psi_{{\eta}_{e2}} ... -\theta_{d } \Psi_{{\eta}_{ed}} + C\Psi_{x} - C_r,
\end{aligned}
\end{equation}
where $\Psi_{x}=H_1 \Psi_{ \zeta_y - \zeta_r} + H_2 \Psi_{ {\zeta}_u}+ C\Psi_{\epsilon}$.
\end{thm}

\begin{pf}
    The proof can be found in Appendices.
\end{pf}

\begin{rem}
Theorem~\ref{thm1} shows that the feedback gain $K^\theta$, computed offline from the nominal system~\eqref{10-eq:supersysw-nom} independently of the exogenous signal $w$ and using the prior datasets $\mathcal{S}_\star$ and $\mathcal{S}_p^\theta$, can exponentially stabilize the subspace $\mathcal{M}$ in~\eqref{18-eq:M}. Note that when $\theta = \theta^\star$, Problem~\ref{problem:1} can be fully solved. Therefore, the next section investigates the online identification of $\theta$ in the exogenous system. 
\end{rem}

\begin{rem}
To relax the excitation condition in Assumption \ref{ass:7}, one may adopt the idea in \cite{Mao} by observing that \eqref{10-eq:supersysw-nom} satisfies the cascaded structure. Moreover, one may also adopt a direct data-driven method by identifying the unknown $H_1,H_2$ in \eqref{10-eq:supersysw-nom} and then computing a stabilizing gain $K^{\theta}$. 
\end{rem}

\subsection{Identifier}

As discussed above, due to the unknown nature of $S$, the exact value of $\theta$ cannot be determined at the initial stage of the system. A natural idea is to generate a regression model of the unknown parameter $\theta^\star$ in the form
\begin{equation*}
\beta(t) = \phi( \theta^{\star}, \alpha(t)),
\end{equation*}
and then identify the parameter online. Unfortunately, we cannot obtain a \emph{linear} regressor in our context, but it can be represented as a linear regressor with some perturbation terms. 
Here, we adopt the discrete-time identifier in \cite{bin2019adaptive}. 



The identifier state is defined as $\xi=(R,v)\in {\mathcal{Z}}:=\mathbb{R}^{d\times d}\times\mathbb{R}^d$, which evolves at jumps according to 
\begin{equation*}
\begin{aligned}
      R^+ & = \mu R + \gamma(\alpha)\gamma(\alpha)^\top
      \\
      v^+ & = \mu v + \gamma(\alpha)\beta
      \\
\theta^+ & \in p_{\mathcal{E}}(R^\dagger v),  
\end{aligned}
\end{equation*}
where $\mu\in(0,1)$ is a forgetting factor, $p_{\mathcal{E}}(\cdot)$ projects onto a compact admissible set ${\mathcal{E}}$, and the reshape operator $\gamma(\cdot)$ has been defined in Section \ref{sec:1} (Notation). Here, we select the variables $(\alpha, \beta)$ as 
\begin{equation}
\label{20-eq:alphabeta}
\alpha = \eta_e, \quad \beta = \theta^\top \eta_e + e.
\end{equation}
The identifier yields the prediction 
$
\hat\beta(\theta,\alpha)=\theta^\top \alpha 
$
with prediction error $\varepsilon(\theta,\alpha,\beta)=\beta-\hat\beta(\theta,\alpha)$ and cost 
$$
J_{\alpha,\beta}(t,j)(\theta)=\sum_{i=0}^{j-1}\mu^{\,j-i-1}\|\varepsilon(\theta,\alpha(t^i,i),\beta(t^i,i))\|^2.
$$
For $(t, j) \in \mathcal{T}$, $t$ denotes the continuous time and $j$ represents the jump iteration index of the discrete identifier. We define
$t^j = \sup_{t \in \mathbb{R}} (t, j) \in \mathcal{T}$, and the construction based on errors $\varepsilon$ and $J$ is intended to identify the optimal linear model that characterizes the relationship between the inputs $\alpha$ and $\beta$. The finite data summaries 
$$
\begin{aligned}
    R^\star(t,j) &= \sum_{i=0}^{j-1}\mu^{\,j-i-1}\gamma(\alpha(t^i,i))\gamma(\alpha(t^i,i))^\top
    \\
v^\star(t,j) &= \sum_{i=0}^{j-1}\mu^{\,j-i-1}\gamma(\alpha(t^i,i))\beta(t^i,i),
\end{aligned}
$$ characterize minimizers of $J_{\alpha,\beta}$ via $R^\star\theta=v^\star$ when $R^\star$ is nonsingular. We provide the necessary clarification of the following property established in \cite{bin2019adaptive}.
\begin{itemize}
    \item[(i)] the recursion for $\xi=(R,v)$ is (robustly) asymptotically stable, under a standard $(J,\varepsilon)$-persistency-of-excitation condition on $\eta_e$,
    i.e., for $\mu \in (0,1)$ and some $\epsilon>0$
\begin{equation*}
    \min ~\sigma \left(
\sum_{i=0}^{j-1} \mu^{j-i-1} \gamma\left( \eta_e(t^i,i) \right)
\gamma\left( \eta_e(t^i,i) \right)^\top
    \right)
    \ge 
    \epsilon.
\end{equation*}
\end{itemize} 
By selecting \eqref{20-eq:alphabeta}, the identifier is interconnected with the proposed internal model such that the identifier updates $\theta$ using the measurable signal $\eta_e$ and the regulation error $e$. According to \cite[Prop. 1]{bin2019adaptive}, the convergence of the identification algorithm can be guaranteed.

\begin{algorithm}[!htp]
\caption{Adaptive Data-Driven Output Regulation}
\label{alg:adaptive_data_driven_OR}
\begin{algorithmic}[1]  
\REQUIRE  Initial parameter $\theta_0,\delta$, dataset $\cals_\star$ for $t\in[0,\tau]$, sample number $N\ge1$, sampling interval $\tau_s=\tau/N$. \COMMENT{$\theta_i$:updated parameter after the $i$-th jump; }
\COMMENT{$\{\cals_{b_i}^{a_i}\}^{N}_{i=0}:=\{(a_i,\, b_i)|\text{ for } i = 0,1,\dots, N\}$; }

\STATE Use $\cals_\star$, \eqref{6-eq:filter}, and \eqref{13-dot:chi} to generate $\tilde \zeta(t), \tilde\chi$:
$$
\begin{aligned}
    \tilde\chi (t) & = \exp(-\Lambda_F t) \mathbf{1}_n,  
    \\
    \tilde \zeta_y(t) & = \int_{0}^{t} \exp[{F(t-\tau)} ] L \tilde y(\tau) \, d\tau, 
    \\
    \tilde \zeta_u(t) & = \int_{0}^{t} \exp[{F(t-\tau)} ] L \tilde u(\tau) \, d\tau,\qquad t \in [0, t_\star ].
\end{aligned}
$$

\ENSURE Boundedness and tracking error $e\to0$.
\FOR{$i=0$ to $N_I$}
\STATE \textbf{Initialization:} $\tilde{\eta}^{\theta_{i}}_e(0)=0\in\mathbb{R}^d$, ${\tilde{\zeta}^{\theta_{i}}}(0)=0\in\mathbb{R}^{2n}$, $\tilde{\chi}^{\theta_{i}}(0)=\mathbf{1}_n$, $\Lambda_F=\diag(\lambda_1,\dots,\lambda_n)$.

\STATE \textbf{Initial offline data of $\theta_i$:} Run the internal model \eqref{9-eq:etay-nom}, i.e.
$$
\tilde \eta_y^{\theta_i}(t) =  \int_{0}^{t} \exp[{\Phi(\theta_i) (t-\tau)} ] G \tilde y(\tau) \, d\tau, \quad t \in[0, t_\star].
$$
Sampling to compute the post-processed data $\cals_p^{\theta_i}$.

\STATE \textbf{Compute Stabilizing Gain $K^{\theta_i}$:} Solve the LMI
\[
\begin{bmatrix}
\hat{\mathcal{A}}^{\theta_i\top} P+P\hat{\mathcal{A}}^{\theta_i}+Q & P\mathcal{B}_c\\
\mathcal{B}_c^\top P & R
\end{bmatrix}\!\preceq\!0,
\quad P=P^\top\!\succ\!0,\ R\!>\!0,
\]
then $K^{\theta_i}=\begin{bmatrix}K_{{\zeta}_y}^{\theta_i}&K_{{\zeta}_u}^{\theta_i}&K_{\eta_e}^{\theta_i}\end{bmatrix}=R^{-1}\mathcal{B}^\top_c P$.

\STATE \textbf{Adaptive Iteration:} 

\STATE Apply $(K^{\theta_{i}},\theta_{i})$ to \eqref{17-filter}; 
\STATE Execute \eqref{17-filter} for duration $T_2$ to obtain $(\eta_e, e)$; update $\theta_{i+1}$ using the discrete identifier.
\IF{$\lvert\theta_{i+1}-\theta_i\rvert<\delta$}  \STATE Set $N_\mathrm{end}:=i$; \textbf{break} \ENDIF
   
\ENDFOR
\RETURN $\{\cals_{e_i}^{\theta_i}\}^{N_\mathrm{end}}_{i=0}$ to generate the image.

\end{algorithmic}
\end{algorithm}

\section{Main Results}
\label{sec:4}
In this section, we present our main results on the data-driven state-feedback controller for output regulation. By integrating the datasets $\cals_\star$ and $\cals_{p}^\theta$, the post-processed internal model, the system tracking filter, and the discrete parameter identifier, we construct the output regulator as follows:
\begin{equation}
\label{21-main_results}
\begin{aligned}
&\left\{
\begin{aligned}
    \dot{\tau} & = 1 
    \\
    (\dot{R}, \dot v) &= (0_{d\times d}, 0_d)
    \\
    \dot{\theta} & = 0 \\
    &\mbox{Eqs.} ~\eqref{17-filter}, ~\eqref{6-eq:filter}, \quad \tau \in [0, T_1+\underline{T}_2] 
            \\
\end{aligned}
\right.
\\
&\left\{
\begin{aligned}
    \tau^+ & = 0 
    \\
    \eta^+_e & = \eta_e 
    \\
    \zeta^+ & = \zeta \qquad (\zeta:=\col(\zeta_y - \zeta_r,\zeta_u))
    \\
    \theta^+ &\in {p_{\mathcal E}}(R^\dagger v) \\
    R^+ &= \mu R + \gamma(\eta_e)\gamma(\eta_e)^\top 
    \\
    v^+ &= \mu v + \gamma(\eta_e)\left(\theta^\top\eta_e + e\right),
    \quad  \tau \in \max[\underline{T}_2, \overline{T}_2],
\end{aligned}
\right.
\end{aligned}
\end{equation}
with the regulator output
\begin{equation*}
    u = K^\theta \begmat{\eta_e \\ \zeta},
\end{equation*}
where $T_1$ denotes the data sampling period of the nominal system, $\underline{T}_2$ represents the lower bound of the steady state convergence time of the actual system, meaning that when $T_2 \ge \underline{T}_2$ the actual system can be guaranteed to have reached steady state, and $\overline{T}_2$ is chosen as the upper bound of the convergence time, with the system output given by \eqref{39-u=kz}. After each jump, the updated parameter $\theta$ and the recorded dataset $\mathcal{S}_\star$ are used to compute the new post-processed dataset $\mathcal{S}_p^\theta$ and the corresponding $K^\theta$.

For the above proposed regulator, we have the following convergence result.
\begin{thm}\label{theorem_last}
Let Assumptions \ref{ass1}--\ref{ass3} and \ref{ass:7} hold for the given plant, the exogeneous system, and the sampling times $t_0, \ldots, t_k$. Then, there exists $T_1^\star>0$ such that for $\underline{T}_2 \ge T_1^\star$, all states in the closed-loop that are interconnected by \eqref{1-eq-sys}, \eqref{2-eq-exosys} and \eqref{21-main_results} are bounded. 
Moreover, suppose that $\eta_e$ is $(J,\epsilon)$-PE for some $J \in \mathbb{N}$ and $\epsilon > 0$, and then there exists $T_2^\star \ge T_1^\star$ such that for any $\underline{T}_2 \ge T_2^\star$  the convergence property in \eqref{eq:e_conv} holds.


\end{thm} 
\begin{pf}
The stability problem of equations \eqref{6-eq:filter} and \eqref{17-filter} in system \eqref{21-main_results} can be addressed by Theorem 8. Considering the existence of the discrete-time identifier, it is necessary to perform a preliminary analysis using Lyapunov inequalities for both continuous-time intervals (i.e., flow processes) and discrete jumps. By analyzing the identification error and the dynamics described by \eqref{19-Phi-CL}, stability can be determined in the jump instants and the minimum iteration time. Additionally, key inequality equations for the boundedness of the identifier states can be derived.
Through the $(J,\epsilon)$-persistency of excitation condition and the Cayley-Hamilton theorem under the true value condition in equation \eqref{19-Phi-CL}, we can extract key information regarding the steady-state regulation error being zero in Problem \ref{problem:1}. The guarantee of asymptotic stability is obtained by integrating the Lyapunov inequalities, both continuous and discrete, that hold for the states of the internal model, identifier, and filter. By combining \cite[Lemma 5]{bin2019adaptive} under the noise-free analysis, we can conclude that $ \lim_{t\to\infty} e(t) = 0.$ The detailed proof can be found in \cite[Proof of Theorem 1]{bin2019adaptive}. Since this paper considers the simplified case without noise, the specific proof details are not repeated here.
\end{pf}
\begin{rem}
The overall algorithm is given in Algorithm \ref{alg:adaptive_data_driven_OR}, whose cascaded structure is illustrated in Fig. \ref{fig:1}.
Note that if the adaptation parameter $\theta$ is equal to their values in $\theta^\star$ and a stabilizing feedback gain $K^{\theta}$ is known, the proposed approach becomes the standard one for model-based output regulation. From this point onward, the convergence is exponential.
\end{rem}

 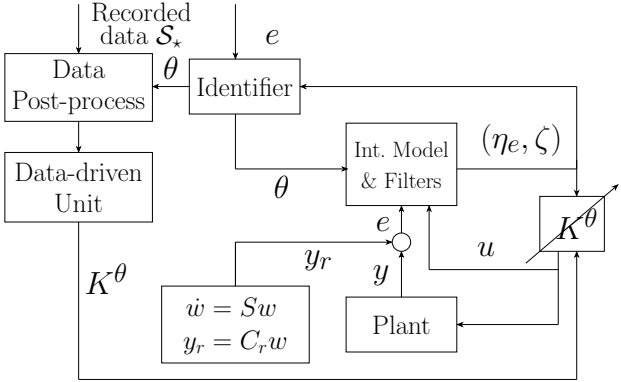
\begin{figure}[!ht]
\centering
\resizebox{.45\textwidth}{!}{%
\begin{circuitikz}
\tikzstyle{every node}=[font=\LARGE]

\draw  (8.75,7.25) rectangle (11.75,5.75);
\node [font=\huge] at (10.25,6.5) {Plant};

\draw  (3.75,7.55) rectangle (7.75,5.45);
\node [font=\huge] at (5.75,6.45) {
$
\begin{aligned}
    \dot w & = Sw
    \\
    y_r & = C_r w
\end{aligned}
$
};

\draw  (8.75,12) rectangle  node {\LARGE 
\begin{minipage}{4cm}
        \centering                   
        Int. Model\\  
        \& Filters
\end{minipage}
} (11.75,9.75);

\draw  (10.25,8.75) circle (0.25cm);

\draw [short] (5.75,7.55) -- (5.75,8.75);
\draw [->, >=Stealth] (5.75,8.75) -- (10,8.75);

\draw [->, >=Stealth] (10.25,7.25) -- (10.25,8.5);
\draw [short] (5.75,7.55) -- (5.75,8.75);
\draw [->, >=Stealth] (10.25,9) -- (10.25,9.75);

\draw  (4.5,13.75) rectangle (7.5,12.25);
\node [font=\huge] at (6,13) {Identifier};
\draw [->, >=Stealth] (5.75,15.25) -- (5.75,13.75);
\node [font=\Huge] at (6.75,14.35) {$e$ };
\draw [short] (5.75,12.25) -- (5.75,10.75);
\draw [->, >=Stealth] (5.75,10.75) -- (8.8,10.75);

\node [font=\huge] at (3.2,15.05) {Recorded };
\node [font=\huge] at (3.2,14.35) {data $\cals_\star$};
\draw [->, >=Stealth] (1.5,15.25) -- (1.5,13.9);

\node [font=\huge] at (1.5,13) {
\begin{minipage}{4cm}
        \centering                   
        {Data}\\  
        Post-process
\end{minipage}
};
\draw  (-0.5,13.9) rectangle (3.5,12.05);

\draw [->, >=Stealth] (4.5,13) -- (3.5,13);

\node [font=\Huge] at (4,13.5) {$\theta$};

\draw  (-0.5,11.2) rectangle (3.5,9.3);
\node [font=\huge] at (1.5,10.25) {
\begin{minipage}{4cm}
        \centering                   
        {Data-driven}\\  
        Unit
\end{minipage}};

\draw [->, >=Stealth] (1.5,12.05) -- (1.5,11.2);

\draw [short] (1.5,9.3) -- (1.5,5);
\node [font=\Huge] at (2.25,7.75) {$K^\theta$};

\node [font=\Huge] at (15,9.25) {$K^\theta$};
\draw  (14,10) rectangle (15.75,8.5);
\draw [short] (14.5,8) -- (11,8);
\draw [->, >=Stealth] (11,8) -- (11,9.75);
\draw [->, >=Stealth] (15,13) -- (7.5,13);
\draw [->, >=Stealth] (14.5,6.5) -- (11.75,6.5);
\draw [->, >=Stealth] (15,11) -- (15,10);
\draw [short] (11.75,10.75) -- (15,10.75);
\draw [short] (15,10.75) -- (15,13);
\draw [short] (14.5,8.5) -- (14.5,6.5);
\draw [->, >=Stealth] (15,5) -- (15,8.5);

\node [font=\Huge] at (7,10.25) {$\theta$};
\node [font=\Huge] at (9.75,9.25) {$e$};
\node [font=\Huge] at (8,8.25) {$y_r$};
\node [font=\Huge] at (9.7,7.85) {$y$};
\node [font=\Huge] at (13.5,11.55) {$(\eta_e,\zeta)$};
\node [font=\Huge] at (12.55,8.55) {$u$};

\draw [->, >=Stealth] (13.6205, 8.1745) -- (16.1295,10.3255);

\draw [short] (1.5,5) -- (15,5);

\end{circuitikz}
}%

    \caption{\raggedright
Block diagram of the closed-loop system structure.
}
\label{fig:1}
\end{figure}

 \section{Numerical Simulations}
\label{sec:5}
In this section, we present numerical simulations to validate the effectiveness of the proposed approach.

We consider a linear system \eqref{1-eq-sys} with the matrices
$$
A=\begin{bmatrix}
1 & 1 & 1\\
-1 & 0 & 1\\
1 & 1 & 0
\end{bmatrix}, 
\;
B = \begin{bmatrix} 0 \\ 1 \\ 2 \end{bmatrix}, 
\;
C=\begin{bmatrix}-1 & 1 & 0\end{bmatrix}.
$$
The reference is generated by the exosystem with $S=[0, -2; 2, 0]$ and $w(0) = \mathbf{1}_2$.
and the regulated output coefficient is set as the column vector $C_r=-[2,{50/\pi^2}]$. The simulation runs for \(t_\star=10\ \mathrm{s}\) with a sampling period \({\tau_s}=0.1\ \mathrm{s}\). The initial state \(x(0)\) is randomly sampled component-wise from the interval \([-1,1]\); the initial conditions \(\epsilon(0)\), \(\eta_e(0)\) and \(\zeta(0)\) are set to zero. The excitation input used for data collection in $\cals_\star$ is $\tilde{u}(t)=\sum_{k=1}^4 k\sin(k\omega_1 t),  \omega_1=5\ \mathrm{rad/s}$. The filter parameters are chosen as $\Lambda_F=-\mathrm{diag}(1,2,3),  ~ {L}=\col(1,2,3).$ The internal model is implemented according to \eqref{6-eq:filter} with initial parameter estimate $\theta_0=\col(1,-1)$ and $G=\col(0,1)$.

For the identifier, we set $\mu=0.9$. At each iteration, $\Phi(\theta)$ is reconstructed and a new stabilizing gain $K^\theta$ is obtained. This process is repeated for $N_I=70$ iterations. The collected data satisfy Assumption \ref{ass:7}. After a number of iteration, the parameter $\theta$ stopped at $\col(-4,0)$ and $K^{\theta_{70}}=-[1.561 \; 1.012 \; 0.348 \; 1.175 \; -1.826 \; 0.714 \; 0.336 \; 1.167].$ We show the simulation results in Fig. \ref{fig:long1} for this configuration.

\begin{figure*}[!t]
     \centering
    \includegraphics[width=0.8\textwidth]{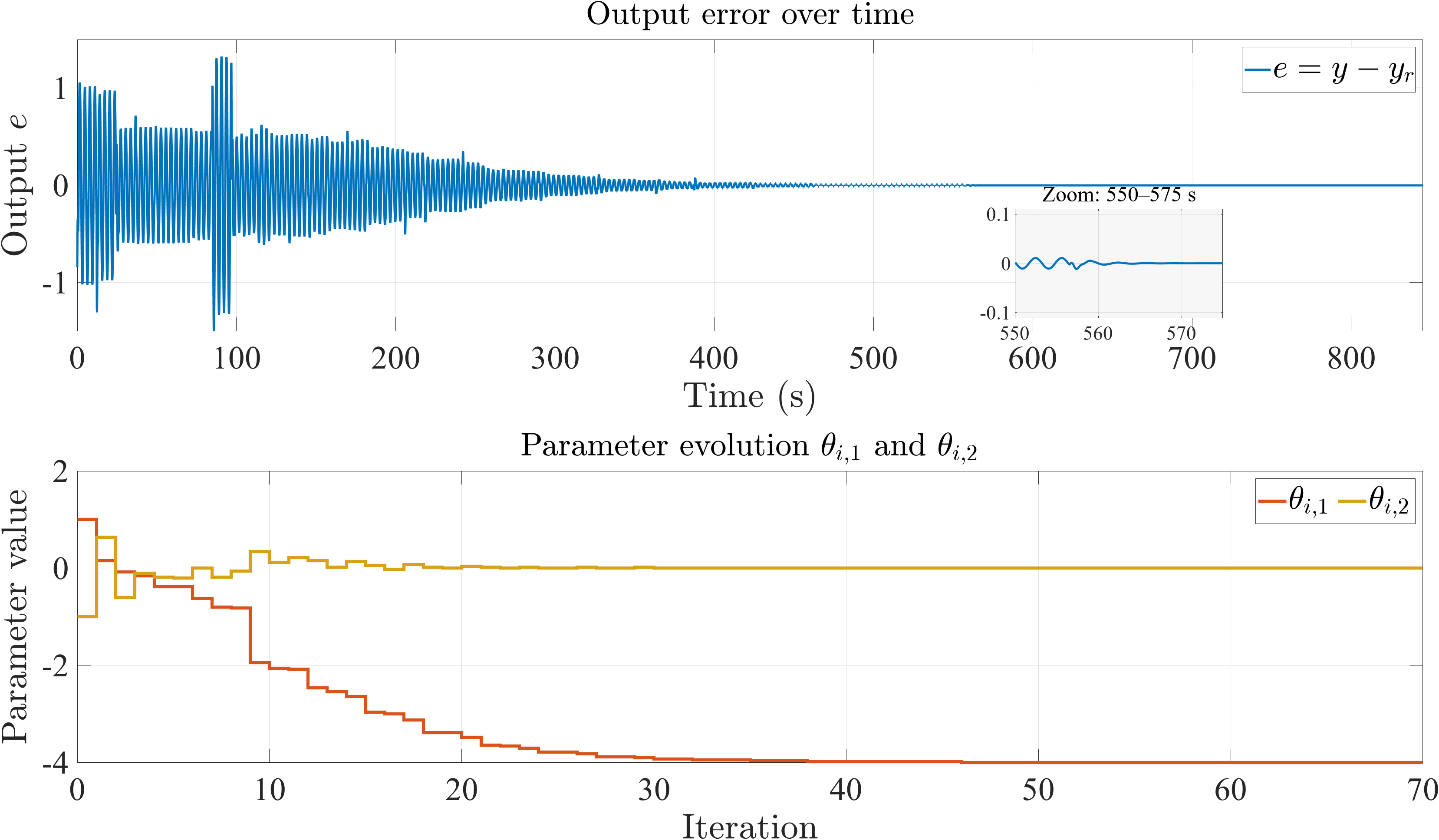}
    \caption{\raggedright
The first figure shows the the regulation error $e$;
the second illustrates the convergence of the parameter $\theta$.
    }
    \label{fig:long1}
\end{figure*}

\section{Conclusion}
\label{sec:6}
\par
In this work, we have developed a data-driven regulator that achieves asymptotic regulation with unknown linear plant and exosystem dynamics.  The nominal system constructed from offline data has been shown to overcome the infeasibility issues caused by the unmeasurable exosystem states, enabling the data-driven computation of stabilizing feedback gain. Moreover, the discrete-time parameter identifier effectively handles the iterative adaptation of the internal-model parameters associated with the unknown exosystem matrix, and the overall scheme has been proven to guarantee stability and asymptotic convergence of the regulation error. Future work will extend this framework to leader–follower multi-agent networks to achieve output synchronization when only a subset of agents has access to the exosystem output. It is also interest to extend the proposed approach to MIMO systems using Kreisselmeier’s adaptive filter~\cite{GAOetal, KRE}, as well as to nonlinear systems through the Kazantzis--Kravaris--Luenberger (KKL) observer~\cite{ANDPRA}.

\appendix

\section{The proof of Lemma \ref{lem8}}
\begin{pf}
For the data-based system equation \eqref{15-datasys}, we separate the known part from the unknown part, leading to
\begin{equation*}
    Z^\theta_+ =  \begmat{\Phi(\theta) & 0  \\
    0 & I_2\otimes F
    }Z^\theta + \begin{bmatrix}
        0 \\ \mathcal{B}
    \end{bmatrix}U + \mathcal{D}\begin{bmatrix}\mathcal{C} & {M}_{\rho} \end{bmatrix}\begin{bmatrix} Z_{\zeta_{oi}} \\ X \end{bmatrix}.
\end{equation*}
Then, the unknown matrix $\mathcal{C}$ can be estimated using measurable data and known matrices.
From Assumption \ref{ass:7}, the following holds, 
\begin{equation*}
  {\mathcal{H}} = \mathcal{D} \mathcal{C} = \left(Z^\theta_+ - \mathcal{A}^{\theta}_n Z^\theta - \mathcal{B}_cU \right)\begin{bmatrix}Z_{\zeta_{oi}}\\  X\end{bmatrix} ^{\dagger} \begin{bmatrix}I_{2n }\\  0_{d\times 2n}\end{bmatrix}.
\end{equation*}
 Hence, we define $\hat{\mathcal{A}}^{\theta}=\mathcal{A}^{\theta}_n + {\mathcal{H}}\begin{bmatrix}0_{2d} & 1 \end{bmatrix},$ so that the matrix $\mathcal{A}^{\theta}_c $ can be expressed as $\hat{\mathcal{A}}^{\theta}$, and is now treated as a known matrix. From Lemma \ref{lem6}, it follows that the pair $(\mathcal{A}^{\theta}_c,\mathcal{B}_c)$ is stabilizable for any fixed $\theta$. 
 

 We now establish closed-loop stability. It is ensured if there exist matrices $P\succ0$ and $Q \succ0$ of proper dimensions and a scalar $R>0$ such that ${\hat{\mathcal{A}}^{\theta\top}} P+P\hat{\mathcal{A}}^{\theta}-P\mathcal{B}_cR^{-1}\mathcal{B}_c^\top P+Q \preceq 0$, with a stabilizing feedback gain $K^{\theta} = R^{-1}\mathcal{B}_c^\top P$. Applying the Schur complement directly leads to \eqref{eq:LMI7} and $\hat{\mathcal{A}}^{\theta} - \mathcal{B}_cK^{\theta}$ is Hurwitz. 
\end{pf}


 

\section{The proof of Theorem \ref{thm1}}
\begin{pf}
To facilitate the proof, we define 
$$
{ \zeta}_e :={ \zeta}_y-{\zeta}_r. 
$$
Its dynamics is given by $\dot{\zeta}_e  = F {\zeta}_e + {Le}$. Define the transverse coordinate 
$$
\rho_e:=  x - \Pi_1  \zeta_{e} -  \Pi_2  \zeta_u,
$$
whose dynamics is $\dot \rho_e =(A - \Pi_1 L C)\rho_e+ \Pi_1LC_rw $. Now, the overall dynamics is 
\begin{equation*}\scalebox{0.91}{$
\begin{aligned}
    \begmat{\dot w \\ \dot \rho_e \\ \dot \eta_e \\ \dot{ \zeta}_e \\ \dot{\zeta}_u}
    = 
\left[\begin{array}{c|cccc}
S & 0 & 0& 0& 0\\
\hline
 \Pi_1LC_r & A - \Pi_1 L C  & 0& 0& 0 \\
-G C_r & GC & \Phi(\theta) & GH_1 & GH_2 
    \\- L C_r & L C  & 0 & F + L H_1 & L H_2
    \\0 &   0& LK^{\theta}_{\eta_e} &  LK^{\theta}_{{\zeta}_e} & F+ LK^{\theta}_{{\zeta}_u}
\end{array}\right] 
    \begmat{w \\ \rho_e \\ \eta_e \\ \zeta_e\\ {\zeta}_u}
\end{aligned}.$}
\end{equation*}
Note that all the sequel analysis is conduct for a fixed $\theta$.
Given the Poisson stability of system $S$ and the Hurwitz property of the $(2,2)$-block above,
the steady state of the overall system can be represented as a linear combination of $w$, i.e.
\begin{equation}\label{B1-EXP}
    \lim_{t\to\infty}
\left|
\begmat{\rho_e(t) \\  \eta_e(t) 
\\
{\zeta}_e (t)
\\
{\zeta}_u(t)}
-
\begmat{
\Psi_{\rho_e} \\ \Psi_{\eta_e} \\ \Psi_{ { \zeta}_e} \\ \Psi_{ {\zeta}_u}
}w(t)
\right| =0 \quad(\mbox{exp.}),
\end{equation}
in which $\Psi_{\rho_e},\Psi_{\eta_e}, \Psi_{ { \zeta}_e}$ and $\Psi_{ {\zeta}_u}$ are solutions of the Sylvester equation 
\begin{equation}\label{B2-sylvst-1}\begin{aligned}
\Psi_{\rho_e}S  &=  (A - \Pi_1 L C)\Psi_{\rho_e} +\Pi_1 LC_r,\\
\Psi_{\eta_e} S  &=  \Phi(\theta)\Psi_{\eta_e}  + GH_1 \Psi_{ {\zeta}_e} +  GH_2 \Psi_{ {\zeta}_u} + GC\Psi_{\rho_e}-G C_r,
\\
\Psi_{ {\zeta}_e} S  &=  (F + L H_1) \Psi_{ {\zeta}_e} +   LH_2 \Psi_{ {\zeta}_u} + LC\Psi_{\rho_e}-L C_r,
\\
\Psi_{ {\zeta}_u} S  &=  LK^{\theta}_{\eta_e}\Psi_{\eta_e}  + LK^{\theta}_{{\zeta}_e} \Psi_{ {\zeta}_e} + ( F+ LK^{\theta}_{{\zeta}_u} )\Psi_{ {\zeta}_u} .
\end{aligned}
\end{equation}

We perform the change of coordinate $(\rho_e, \eta_e, \zeta_e, \zeta_u) \mapsto \mathbf{x}:= (\rho_e - \Psi_{\rho_e}w , \eta_e - \Psi_{\eta_e}w, \zeta_e - \Psi_{\zeta_e}w, \zeta_u - \Psi_{\zeta_u}w )$, and we have
\begin{equation}\label{B3-X}\begin{aligned}\dot{\bf x} =
   \begin{bmatrix}
  A-\Pi_1 L C  & 0& 0& 0 \\
 GC & \Phi(\theta) & GH_1 & GH_2 
    \\ L C  & 0 & F + L H_1 & L H_2
    \\   0& LK^{\theta}_{\eta_e} &  LK^{\theta}_{{\zeta}_e} & F+ LK^{\theta}_{{\zeta}_u}
\end{bmatrix}{\bf x}.\end{aligned}
\end{equation}
Considering that \eqref{B3-X} satisfies the stabilizability condition in Lemma \ref{lem6} and that Lemma \ref{lem8} guarantees the solvability of $K^{\theta}$, a system matrix in \eqref{B3-X} can be made Hurwitz. Consequently, the exponential convergence property of \eqref{B1-EXP} follows. If we partition conformally $\Psi_{\eta_e}:= \col(\Psi_{\eta_{e1}}, \ldots, \Psi_{\eta_{ed}})$, then \eqref{B2-sylvst-1} becomes \eqref{19-Phi-CL}, where $C\Psi_{x} - C_r= H_1 \Psi_{ {\zeta}_e} + H_2 \Psi_{ {\zeta}_u} + C\Psi_{\rho_e} - C_r.$ By substituting the first $d-1$ equations into the last equation and noting the transformation relation $ \Psi_x=\Psi_{\rho_e}+ \Pi_1  \zeta_e +  \Pi_2  \zeta_u  $, we obtain $\Psi_{{\eta}_{ed}} S =  -\theta_1 \Psi_{{\eta}_{e1}} -\theta_2 \Psi_{{\eta}_{e2}} ... -\theta_{d } \Psi_{{\eta}_{ed}} + C\Psi_{x} - C_r$, which yields the desired identity and completes the proof.
\end{pf}

\begin{figure}
    \centering

\end{figure}

\bibliography{ifacconf}             
                                                   







\end{document}